\documentclass[a4paper,12pt]{article} 

\usepackage{amsmath}
\usepackage{color,graphicx}
\usepackage[top=2.5cm,right=2cm,bottom=2.5cm,left=2cm]{geometry}
\usepackage[colorlinks,linkcolor=blue,citecolor=blue,pagebackref=false]{hyperref}
\usepackage{cite}
\usepackage[parfill]{parskip}
\usepackage{caption}
\usepackage{subcaption}
\usepackage{float}
\usepackage{hyperref}
\usepackage{mathtools}
\usepackage{eurosym}

\makeatletter
\newcommand{\rightleftarrows}[2]{%
\mathrel{\mathop{%
  \vcenter{\offinterlineskip\m@th
    \ialign{\hfil##\hfil\cr
      \hphantom{$\scriptstyle\mspace{8mu}{#1}\mspace{8mu}$}\cr
        \rightarrowfill\cr
        \vrule height0pt width 2em\cr
        \leftarrowfill\cr
        \hphantom{$\scriptstyle\mspace{8mu}{#2}\mspace{8mu}$}\cr
        \noalign{\kern-0.3ex}
        }%
      }%
    }\limits^{#1}_{#2}}%
  }
\makeatother

\begin{document}

\begin{center}
\textbf{A theoretical framework to explain\\ non-Nash equilibrium strategic behavior in experimental games}

\vspace{0.5 cm}

Mojtaba Madadi Asl\textsuperscript{1*}\let\thefootnote\relax\footnotetext{* Corresponding author. Email address: m.madadi@ipm.ir.}, and
Mehdi Sadeghi\textsuperscript{2}
\\
\bigskip
\footnotesize{
\textsuperscript{1}School of Biological Sciences,\\Institute for Research in Fundamental Sciences (IPM), Tehran, Iran\\
\textsuperscript{2}Department of Medical Genetics,\\National Institute of Genetic Engineering and Biotechnology (NIGEB), Tehran, Iran
}

\end{center}


\vspace{0.7cm}

\begin{center}
{\small
\begin{minipage}{14cm}

\begin{center}
\textbf{Abstract}
\end{center}

\vspace*{0.2cm}

Conventional game theory assumes that players are perfectly rational. In a realistic situation, however, players are rarely perfectly rational. This bounded rationality is one of the main reasons why the predictions of Nash equilibrium in normative game theory often diverge from human behavior in real experiments. Motivated by the Boltzmann weight formalism, here we present a theoretical framework to predict the non-Nash equilibrium probabilities of possible outcomes in strategic games by focusing on the differences in expected payoffs of players rather than traditional utility metrics. In this model, bounded rationality is parameterized by assigning a temperature to each player, reflecting their level of rationality by interpolating between two decision-making regimes, i.e., utility maximization and equiprobable choices. Our framework predicts all possible joint strategies and is able to determine the relative probabilities for multiple pure or mixed strategy equilibria. To validate model predictions, by analyzing experimental data we demonstrated that our model can successfully explain non-Nash equilibrium strategic behavior in experimental games. Our approach reinterprets the concept of temperature in game theory, leveraging the development of theoretical frameworks to bridge the gap between the predictions of normative game theory and the results of behavioral experiments.\\


\end{minipage}
}
\end{center}

\vspace{0.5cm}



\section{Introduction}

The theory of rational choice assumes that each player chooses the best available strategy to maximize her expected utility in an interactive manner, depending on the choice of other players~\cite{boudon2003beyond}. However, the assumption of perfect rationality rarely holds in a realistic situation, e.g., due to noise in environment, information-processing costs or innate, psychological features of the subjects, whether the players
are humans, animals, or computational agents~\cite{wolpert2006information,ortega2013thermodynamics}. This bounded rationality is one of the major reasons why the predictions of normative game theory (i.e., what players are rationally required - or ought - to do) often diverge from human behavior in real experiments (i.e., what players actually do)~\cite{simon1956rational,simon1972theories}, e.g., in the context of economics and psychology as well as social and cognitive sciences~\cite{camerer2003behavioural,camerer2006does,camerer2011behavioral,camerer2015psychological,rhoads2024modelling}.

In fact, human behavior does not always conform to the assumptions of perfect rationality and self-interest that underlie normative game theory. Instead, human decisions are influenced by social, psychological, cultural and emotional factors that are not fully captured by the rational player model. For example, the results of the Ultimatum game has been the subject of debate in experimental economics~\cite{guth1982experimental}. The Ultimatum game is a simple two-player bargaining game where one player (the proposer) is granted an amount of money, say \EUR10 (in the units of \EUR1). She offers any part of it (say, \EUR$x$) to a second player (the responder), which can either accept or reject the offer. If she accepts, the money is shared as proposed (i.e., \EUR$x$ and \EUR$10-x$), and if she rejects, then neither of the players receives anything. In its simplest version, in either case the game ends. The only subgame-perfect Nash equilibrium is to offer an infinitesimal amount (here, \EUR1) and to accept it. However, such a theoretically predicted equilibrium is inconsistent with the findings of standard experiments indicating that the mean accepted offers in industrialized societies are typically around 40-50\% of the endowed amount~\cite{roth1991bargaining,henrich2005economic}.

Another well-known example is the Keynsian beauty contest~\cite{keynes1937general}, later recognized as the Guess $\frac{2}{3}$ of the Average game~\cite{ledoux1981concours}. In this game, participants are asked to choose an integer number (e.g., in the range 0-100) as close as possible to $\frac{2}{3}$ of the average of all guesses, where the winner is the one who guessed the closest number to $\frac{2}{3}$ of the average guess. Equilibrium analysis reveals that this game has a unique pure strategy Nash equilibrium (i.e., the winning guess is 0) which can be found by iterated elimination of weakly dominated strategies. In order for Nash Equilibrium to be realized, players must assume that all other players are rational and that there is common knowledge of rationality. But, the winning value was found to be around 20 in experiments among people~\cite{nagel1995unraveling}, suggesting that people are not perfectly rational and do not assume common knowledge of rationality.

To account for such discrepancies between theory and experiment, mathematical frameworks were developed to predict human behavior in strategic situations~\cite{camerer2004behavioral}. Unlike the Nash equilibrium which assumes perfect rationality, non-Nash equilibrium concepts such as the quantal response equilibrium~\cite{mckelvey1995quantal,mckelvey1998quantal}, epsilon equilibrium~\cite{radner1980collusive} and level-k reasoning or cognitive hierarchy models~\cite{nagel1995unraveling,camerer2004cognitive,costa2006cognition,evans2024bounded} were developed to model bounded rationality. Quantal response models usually use the Boltzmann distribution to approximate choice probabilities of players whose rationality is influenced by a \textit{temperature} parameter~\cite{mckelvey1995quantal,mckelvey1998quantal}. On the other hand, the level-k model usually assumes that players are characterized by various depth of reasoning~\cite{nagel1995unraveling,camerer2004cognitive}, allowing them to approach the game naively (a level-0 player) or perfectly rational (a level-k player), where k represents the number of reasoning iterations. However, iterated reasoning in humans is constrained by a limited cognitive capacity~\cite{camerer2004behavioral}.

A number of previous studies applied thermodynamics and statistical physics to the problem of bounded rational decision-making by employing the Boltzmann distribution~\cite{mckelvey1995quantal,mckelvey1998quantal}. This approach draws an analogy between energy and utility, i.e., just as physical systems tend to select states characterized by the lower energies, decision makers would like to choose states with greater utility~\cite{babajanyan2020energy}. Perfect rationality is then analogous to physical systems with zero temperature, where the lowest energy state is the most probable outcome. Stochastic choice paradigms such as logit choice models were inspired by the Boltzmann weight formalism and have been widely explored in interdisciplinary physics~\cite{blume1993statistical,helbing1993stochastic,ortega2013thermodynamics,nakamura2019cooperation}.

Yet, these models cannot predict all possible joint mixed strategies chosen by the players as well as the relative probabilities of the joint strategies. Furthermore, they lack the interpretation of playing probabilistically even when the players are perfectly rational, failing to produce a full distribution over all possible outcomes in a game. To address these issues, here we argue that assigning a temperature to each player can serve as a quantitative measure to approximate their sanity in the game such that cold players decide rationally, whereas hot player choose rather irrationally. Here, temperature is a concept to decribe the level of uncertainty or random decisions in players' strategies. Inspired by the Boltzmann weight formalism borrowed from statistical physics, we present a theoretical framework which is able to predict non-Nash equilibrium probabilities of possible outcomes in two-person strategic games. Temperature variations allow the players to interpolate between two different decision-making regimes, i.e., utility maximization (representing decisions made by perfectly rational players at zero temperature) and equiprobable choices (representing decisions made by perfectly irrational players at infinite temperature). We show that relative probabilities of the joint strategies of the players can be determined by minimizing the difference between the expected payoffs of players attributed to each set of probability distributions with respect to the Nash equilibrium of the game, weighted by the Boltzmann function.

Our theoretical framework provides a full distribution over all possible outcomes in a game as well as their probabilities by predicting all possible joint pure or mixed strategies. The core of our model relies on highlighting the importance of the expected payoff difference of the players' strategies rather than the traditional utility metrics, allowing for the prediction of players' temperature and the relative probability of their choices in a non-Nash equilibrium state. Whether players employ pure or mixed strategies, our framework provides a full description over all possible joint strategies and is able to determine the relative probabilities for different choices as well as for multiple pure or mixed strategy Nash equilibria.

To validate model predictions, by analyzing experimental data we demonstrated that our model can successfully explain non-Nash equilibrium strategic behavior in the experimental Dictator game and Ultimatum game. To that end, first we extracted the temperature of players based on the amounts of points they offered in each experimental game. Then, we used the rationality temperature extracted from the cohort playing Dictator game to predict the behavior of players in the Ultimatum game and vice versa. Our results show that the distributions of offers predicted by the model are in good agreement with the experimentally observed offers in each game. Our approach offers a generic solution concept for predicting non-Nash equilibrium probabilities in two-person games based on the temperature of players and may account for the discrepancy between the predictions of normative game theory and the results of behavioral experiments.


\section{Methods}
\subsection{Theoretical framework}
\subsubsection{Transition probability between two strategies}

Assume a generic two-player strategic game where each player is confronted with a binary choice, i.e., she can choose between two possible pure strategies $\mathcal{S} = \lbrace s_1,s_2 \rbrace$ with ordinal preferences characterized by the utility (payoff) function $\mathcal{U}(\mathcal{S}) = \lbrace u(s_1,s_1),u(s_1,s_2),u(s_2,s_1),u(s_2,s_2) \rbrace$; an optimal quantity that a rational player aims to maximize. Each $u(s_i,s_j)$ pair represents the payoff to player 1, for example, when she chooses the strategy $s_i$ and player 2 chooses the strategy $s_j$. Instead of deterministic choices, assume that each player makes her decisions stochastically with a probability, e.g., due to her imperfect rationality. Therefore, a player can switch between two possible states (i.e., two strategies) in time, as follows:
\begin{equation}\label{eq:1}
s_1 \rightleftarrows{~~r_{12}~~}{~~r_{21}~~} s_2,
\end{equation}
where $r_{12}$ ($r_{21}$) is the rate at which a player switches from $s_1 \rightarrow s_2$ ($s_2 \rightarrow s_1$). Without loss of generality and for analytical tractability, we assume that these transitions occur uniformly and randomly at a constant rate, meaning that the players have no memory of past interactions. This is motivated by the fact that here we focus on strategic games lacking a temporal dimension in which actions are chosen once and for all (i.e., a one-shot game). Therefore, the probability that a player jumps from $s_1 \rightarrow s_2$ ($s_2 \rightarrow s_1$) in the time interval $(t , t + dt)$ can be represented by $r_{12} \, dt$ ($r_{21} \, dt$).

Assuming that the dynamics of the system started at some initial time $t_0$, the probabilities $p(s_1,t)$ and $p(s_2,t)$ that a player may choose the strategy $s_1$ and $s_2$ at time $t$, respectively, is given by the following master equations:
\begin{equation}\label{eq:2}
\begin{aligned}
\dfrac{dp(s_1,t)}{dt} &= - r_{12} \, p(s_1,t) + r_{21} \, p(s_2,t), \vspace{0.15cm} \\
\dfrac{dp(s_2,t)}{dt} &= - r_{21} \, p(s_2,t) + r_{12} \, p(s_1,t),
\end{aligned}
\end{equation}
where the sum of the probabilities satisfy the condition $p(s_1,t) + p(s_2,t) = 1$. For simplicity, the terms of order $\mathcal{O}(dt^2)$ were ignored. In the derivation of Eq.~(\ref{eq:2}), we used the first-order Markov property which assumes that transition from one state to the other during the time interval $(t , t + dt)$ only depends on the current state at time $t$, not on the prior history of the system~\cite{gagniuc2017markov}. Therefore, Eq.~(\ref{eq:2}) represents a relation for the probability that a stochastic player which can switch between two strategies is in one of these states at time $t$.

The relation $\frac{d}{dt} [p(s_1,t) + p(s_2,t)] = 0$ implies that $p(s_1,t) + p(s_2,t) = 1$ is valid at all times provided that the initial condition satisfies $p(s_1,t_0) + p(s_2,t_0) = 1$. Assuming that the transition rates are constant over time (time homogeneity)~\cite{gagniuc2017markov}, Eq.~(\ref{eq:2}) can be solved as follows:
\begin{equation}\label{eq:3}
\begin{aligned}
p(s_1,t) &= p(s_1,t_0) \,  \dfrac{r_{21} + r_{12} e^{-(r_{12} + r_{21})(t - t_0)}}{r_{12} + r_{21}} + p(s_2,t_0) \,  \dfrac{r_{21} (1 - e^{-(r_{12} + r_{21})(t - t_0)})}{r_{12} + r_{21}}, \vspace{0.25cm} \\
p(s_2,t) &= p(s_1,t_0) \, \dfrac{r_{12} (1 - e^{-(r_{12} + r_{21})(t - t_0)})}{r_{12} + r_{21}} + p(s_2,t_0) \, \dfrac{r_{12} + r_{21} e^{-(r_{12} + r_{21})(t - t_0)}}{r_{12} + r_{21}},
\end{aligned}
\end{equation}
where the steady-state solution of Eq.~(\ref{eq:3}) in the limit $t \rightarrow \infty$ can be derived by using the normalization condition $p(s_1,t_0) + p(s_2,t_0) = 1$, as follows:
\begin{equation}\label{eq:4}
\begin{aligned}
p(s_1) &= \dfrac{r_{21}}{r_{12} + r_{21}}, \vspace{0.25cm} \\
p(s_2) &= \dfrac{r_{12}}{r_{12} + r_{21}},
\end{aligned}
\end{equation}
which implies that in a game that each player has two strategies the stationary distributions satisfy the detailed balance condition~\cite{tauber2014critical}, i.e., $r_{12} \, p(s_1) = r_{21} \, p(s_2)$.


\subsubsection{Logit transition rates and Boltzmann weight}

In general, detailed balance considerations~\cite{tauber2014critical} imply that the rates with which the players update their decisions are such that at long times ($t \rightarrow \infty$) the equilibrium Boltzmann probability distribution is recovered~\cite{anderson1992discrete,blume1993statistical,bouchaud2013crises}:
\begin{equation}\label{eq:5}
p(s_i) = \frac{e^{\beta u(s_i)}}{\sum_j e^{\beta u(s_j)}},
\end{equation}
where $u(s_i)$ is the utility of alternative $s_i$, and $\beta$ represents the rationality of each player such that in the limit $\beta \rightarrow 0$ the player is assumed to be perfectly irrational, whereas $\beta \rightarrow \infty$ indicates perfect rationality of the player~\cite{wolpert2006information}. The summation over all possible strategies is a normalization factor such that $\sum_i p(s_i) = 1$. One way to interpret $p(s_i)$ is that stochasticity in player's choice can be linked to different aspects such as an imperfect knowledge of the utilities, imprecise perception of the utilities, cognitive limitations, context or environment which are all included in the rationality level $\beta$. Assuming that each player switches between different alternatives, $p(s_i)$ makes sense even for a single player, i.e., it is the probability that, at a given instant of time she chooses the alternative $s_i$~\cite{bouchaud2013crises}. The notion behind the exponential form of probability given in Eq.~(\ref{eq:5}) was previously addressed in a number of studies~\cite{anderson1992discrete,nadal1998formal,wolpert2006information,bouchaud2013crises} by using axiomatic arguments, or based on a randomized interpretation of the perceived utilities, or based on entropy arguments. Accordingly, the ratio of transition rates between the two arbitrary states can be written by using the Boltzmann factor~\cite{bouchaud2013crises}:
\begin{equation}\label{eq:6}
\dfrac{r_{12}}{r_{21}} = e^{\beta \Delta U_{21}},
\end{equation}
where $\Delta U_{12} = u(s_2) - u(s_1)$ ($\Delta U_{21} = u(s_1) - u(s_2)$) is the player's utility difference when she switches from $s_1 \rightarrow s_2$ ($s_2 \rightarrow s_1$). At $\beta \rightarrow 0$ (i.e., a perfectly irrational player), the ratio of transition rates approaches unity ($\frac{r_{12}}{r_{21}} \rightarrow 1$) indicating that the transition rate is extremely fast and the player make her decisions with a 50-50$\%$ chance on average, such that $p(s_1) = p(s_2) = \frac{1}{2}$. On the other hand, at $\beta \rightarrow \infty$ (i.e., a perfectly rational player) the ratio of transition rates approaches zero or infinity ($\frac{r_{12}}{r_{21}} \rightarrow 0$ such that $p(s_1) = 0$, $p(s_2) = 1$ or $\frac{r_{12}}{r_{21}} \rightarrow \infty$ such that $p(s_1) = 1$, $p(s_2) = 0$). This implies that the transition rates are frozen at zero or unity, i.e., $(r_{12},r_{12}) = (0,1)$ or $(r_{12},r_{12}) = (1,0)$, depending on the utility difference for each transition ($\Delta U > 0$ or $\Delta U < 0$).

There are several conventional approaches to formulate transition rates in Eqs.~(\ref{eq:4}) and (\ref{eq:6})~\cite{holehouse2022non}. A standard choice is to consider Glauber dynamics or logit transition rates~\cite{glauber1963time,blume1993statistical,mckelvey1995quantal,nadal1998formal,bouchaud2013crises}:
\begin{equation}\label{eq:7}
\begin{aligned}
r_{12} &= \dfrac{1}{1 + e^{- \beta \Delta U_{12}}}, \vspace{0.20cm} \\
r_{21} &= \dfrac{1}{1 + e^{- \beta \Delta U_{21}}}.\\
\end{aligned}
\end{equation}

The logit transition rates introduced in Eq.~(\ref{eq:7}) were previously derived using the maximum entropy principle~\cite{nadal1998formal,wolpert2006information}. This assumption suggests that players make their decisions by balancing their immediate gains in terms of the utility of the selected strategy with a curiosity to explore alternative options based on their level of rationality in the game.


\subsubsection{Relative entropy}

Relative entropy, also known as Kullback-Leibler (KL) distance, quantifies the difference between two probability distributions. It is always non-negative, and is zero if and only if the probability distributions are identical. Relative entropy can be used to calculate the effective information which measures the uncertainty reduction provided by an \textit{a posteriori} repertoire (observed state of the system) with respect to an \textit{a priori} repertoire (maximum entropy distribution on the states of the system), as follows~\cite{balduzzi2008integrated}:
\begin{equation}\label{eq:8}
I(\mathcal{P},\mathcal{P}^{\prime}) = H_{\rm max}(\mathcal{P}) - H_{\rm obs}(\mathcal{P}^{\prime}),
\end{equation}
where $I$ is the effective information, $\mathcal{P} = \lbrace p_1,p_2,...,p_n \rbrace$ and $\mathcal{P}^{\prime} = \lbrace p^{\prime}_1,p^{\prime}_2,...,p^{\prime}_n \rbrace$ are the probability distributions over possible outcomes, and $H_{\rm max}(\mathcal{P})$ and $H_{\rm obs}(\mathcal{P}^{\prime})$ are the corresponding maximum and observed Shannon entropies, respectively. The Shannon entropy is defined as the entropy of a probability distribution, e.g., over the possible outcomes of a game:
\begin{equation}\label{eq:9}
H(\mathcal{P}) = - \sum_i p_i \log(p_i),
\end{equation}
where $\log(p)$ denotes the logarithm in base 2 such that entropy gives the unit of bits.


\subsection{Experimental paradigm}

We used previously published~\cite{wu2019gossip} experimental data, where participants were involved in two different two-person games, i.e., a Dictator game or an Ultimatum game. The experimental dataset analyzed during this study is publicly available at \href{https://osf.io/ft8mu/}{https://osf.io/ft8mu/}.


\subsubsection{Participants, design and procedure}

As indicated in the original study~\cite{wu2019gossip}, 240 students (137 female, mean age $21.53 \pm 3.44$ years) were recruited from a Dutch university to participate in the experiment. The experiment was composed of an initial game (i.e., Dictator game or Ultimatum game) followed by a Trust game with gossip for interaction between individuals. The participants received a baseline payment of \EUR2 and could earn an extra bonus of up to \EUR4 depending on their own and other players' decisions. The experiment was conducted in a psychology lab. Data were collected across 39 sessions with 6 or 12 participants in each session. During the experiment, participants were engaged in two decision-making tasks: (i) An initial game comprising a Dictator game or an Ultimatum game involving person A and person B, and (ii) a subsequent Trust game involving person C and person A~\cite{wu2019gossip}. The participants were randomly assigned as person A, B, or C. Thus, the 240 participants were grouped into 80 triads including person A, B, and C, with 20 triads in each condition. Of note, in this analysis, we only considered data pertaining to the initial game in the experiment, specifically describing the outcome of a Dictator game or an Ultimatum game between the two players, i.e., data relating to person A and person B (referred to as player 1 and player 2 from now on).


\subsubsection{Dictator game and Ultimatum game}

In the experiment~\cite{wu2019gossip}, the Dictator game involved two players, i.e., an allocator (player 1) and a receiver (player 2). The allocator was initially endowed with 100 monetary points (= \EUR1) and could allocate arbitrary points in the range 0-100 between herself and the receiver, while the receiver had no choice but to accept the allocated amount. Both players were informed about the final outcome after the allocator made the decision. The Ultimatum game involved two players, i.e., a proposer (player 1) and a responder (player 2). The proposer was initially endowed with 100 points (= \EUR1) and could offer arbitrary points in the range 0-100 to the responder, who could then accept or reject the offer. If the offer was accepted, the 100 points would be shared as proposed, otherwise, neither of the players received anything. Both players were informed about the proposed offer and the final outcome after their respective decisions, whether the offer was accepted or rejected.


\section{Results}
\subsection{Theoretical framework}
\subsubsection{Rationality as a thermometer for players' sanity}

As discussed in the Methods, Eqs.~(\ref{eq:4})-(\ref{eq:7}) provide a framework that allows for a probabilistic estimation of the players' choices based on their rationality ($\beta$). In this context, rationality has been interpreted as an analogue of the inverse temperature ($\beta \equiv \frac{1}{T}$) widely used in thermodynamics and statistical physics~\cite{bianconi2001bose}. In game theory, the notion of temperature is used in a metaphorical sense rather than a literal thermodynamic temperature. In this mindset, temperature is simply a conceptual tool to describe the level of randomness or uncertainty in players' strategies. A cold player with lower temperature decides more deterministically in a way that maximizes her utility gain, whereas a hot player with higher temperature decides more randomly. Therefore, perfect rationality corresponds to $T \rightarrow 0$ where players make their decisions so that their utility is maximized, whereas perfect irrationality corresponds to $T \rightarrow \infty$ where players make their choices randomly with a uniform probability.


\subsubsection{Temperature concept and Nash equilibrium}

A strategic game is a model of interacting players where each player has a set of $n$-dimensional possible pure strategies $\mathcal{S} = \lbrace s_1,s_2,...,s_n \rbrace$ with a utility function characterizing the player's payoff, $\mathcal{U}(\mathcal{S}) = \lbrace u(s_1,s_1,...,s_1),...,u(s_n,s_n,...,s_n) \rbrace$, which reflects the player's ordinal preferences with respect to the choices of other players, such that each argument in $u(.)$ corresponds to a strategy chosen by each player in the game, i.e., player 1, player 2, and so on. For example, in $u(s_1,s_1,...,s_1)$ all players have chosen the strategy $s_1$. As in the theory of rational choice, we assume that each player chooses the best available strategy according to her utility function, depending on the strategies of the other players. Nash equilibrium of such a strategic game is a deterministic strategy profile in which every player's strategy is optimal given every other player's strategy, corresponding to a steady-state solution of an idealized situation where no player wishes to deviate~\cite{nash1950equilibrium}. The notion of Nash equilibrium can be generalized to model stochastic steady state of a strategic game with von Neumann-Morgenstern (vNM) preferences where each player is allowed to assign a probability distribution $\mathcal{P}(\mathcal{S}) = \lbrace p(s_1),p(s_2),...,p(s_n) \rbrace$ to her set of strategies (i.e., a mixed strategy) such that $\sum_i p(s_i) = 1$~\cite{von2007theory}, rather than a restriction to choose deterministically (i.e., a pure strategy). In this case, for example, $p(s_1)$ is the probability assigned by the player's mixed strategy $p$ to her strategy $s_1$.

In this setting, the players' preferences with respect to lotteries over strategies are represented by the expected value of a payoff function over strategies. By definition, the mixed strategy profile $p^*$ is a Nash equilibrium of a strategic game with vNM preferences if for every player $i$ with every mixed strategy $p_i$, the expected payoff to the player $i$ of $p^*$ is at least as good as the expected payoff to the player $i$ of $(p_i,p^*_{-i})$, satisfying~\cite{osborne2004introduction}:
\begin{equation}\label{eq:10}
\mathcal{E}_i(p^*) \geq \mathcal{E}_i(p_i,p^*_{-i}),
\end{equation}
where $\mathcal{E}_i(p) = \sum_{s_i \in \mathcal{S}} p_i(s_i) \mathcal{E}_i(s_i,p_{-i})$ represents the player $i$'s expected payoff to the mixed strategy $p$, such that $\mathcal{E}_i(s_i,p_{-i})$ is her expected payoff when she uses the pure strategy that assigns probability 1 to $s_i$ and every other player $j$ uses her mixed strategy $p_j$, weighted by the probability $p_i(s_i)$ assigned to the strategy $s_i$ by player $i$'s mixed strategy $p_i$, and the subscript $-i$ stands for except $i$. In special cases, a mixed strategy allows for the possibility of assigning probability 1 to a single strategy and 0 to others (i.e., a pure strategy).

Now, assume that every player $i$ is characterized by a temperature $T_i$ which quantifies her level of rationality. A distribution of the players' temperature where all players do not share the same level of rationality results in a probability distribution over the possible strategies for each player which is not necessarily a mixed strategy Nash equilibrium. Here, we argue that pure or mixed strategy Nash equilibrium corresponds to $T_i \rightarrow 0$ for all players, indicating that they are perfectly rational. The question then arises: What strategies with what probabilities will be chosen by the players at other temperatures and what is the steady-state solution of such a strategic game?

To address these questions, let us represent a few classical examples. For simplicity, consider a two-person two-strategy ($2 \times 2$) strategic game, such as one of those shown in Table~\ref{table1}. In each game, the players are characterized by strategies $\mathcal{S} = \lbrace s_1,s_2 \rbrace$ and utility function $\mathcal{U}(\mathcal{S}) = \lbrace u(s_1,s_1),u(s_1,s_2),u(s_2,s_1),u(s_2,s_2) \rbrace$. The two rows (columns) in Table~\ref{table1} correspond to the two possible strategies of player 1 (player 2), where the best response of each player is marked with an asterisk, i.e, any strategy $s_i$ for the player $i$ that is at least as good as every other strategies of the player $i$ when the other players' strategies are given by $s_{-i}$. Of note, we sometimes simplify the notation by assuming that the individual in the role of player 1 plays strategies $s_1$ and $s_2$ with probabilities $p_1(s_1) = p$ and $p_1(s_2) = 1 - p$, respectively. The probability that player 2 plays strategies $s_1$ and $s_2$ are $p_2(s_1) = q$ and $p_2(s_2) = 1-q$, respectively. In either case, the sum of probabilities for every player $i$ satisfies $p_i(s_1) + p_i(s_2) = 1$.

\begin{figure}[t!]
\centering
\includegraphics[scale = 1]{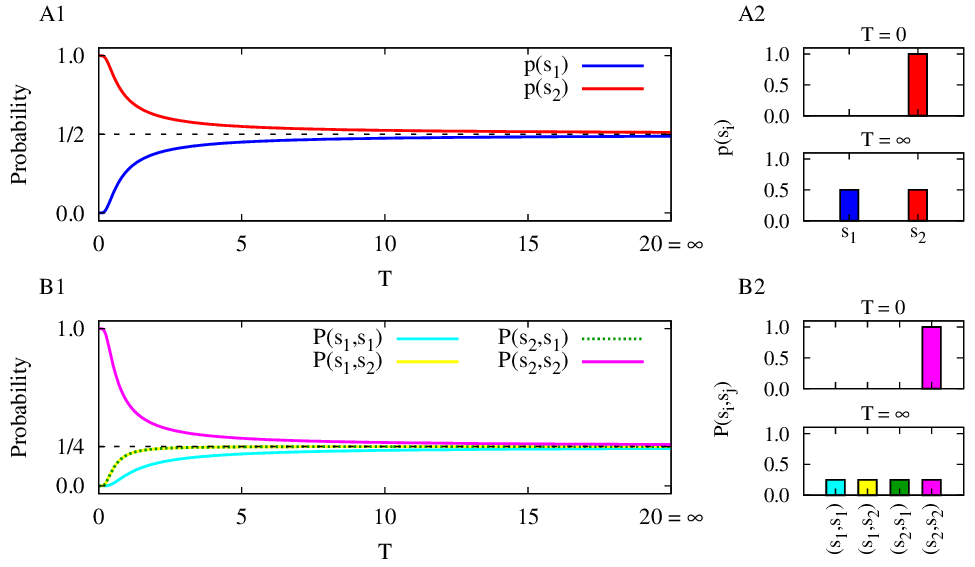}
\caption{{\bf Temperature-based probabilities of possible outcomes of the Prisoner's Dilemma.} (\textbf{A1}) Probability distributions that a player chooses the strategy $s_1 = Quiet$ or $s_2 = Betray$ based on her temperature (rationality) assuming that $T_1 = T_2 = T$. (\textbf{A2}) Probability of choosing each strategy when the player is perfectly rational ($T \rightarrow 0$, top) or perfectly irrational ($T \rightarrow \infty$, bottom). (\textbf{B1}) Joint probabilities of possible outcomes of the game. (\textbf{B2}) Joint probability of possible outcomes of the game when the two players are perfectly rational ($T \rightarrow 0$, top) or perfectly irrational ($T \rightarrow \infty$, bottom). In the figure, temperature is dimensionless and $T = 20$ was sufficient to be considered as $T \rightarrow \infty$ where probabilities approach their asymptotic values. Dashed horizontal lines represent asymptotic probabilities at $T \rightarrow \infty$ as a reference.}
\label{fig1}
\end{figure}

For starters, consider the Prisoner's Dilemma whose payoff table is shown in Table~\ref{table1} (left). The players are faced with a binary choice such that $\mathcal{S} = \lbrace s_1,s_2 \rbrace = \lbrace Quiet,Betray \rbrace$. A quick inspection of the four possible pairs of strategies in Table~\ref{table1} (left) reveals that Prisoner's Dilemma in this form does not have a stochastic steady state, i.e., a mixed strategy Nash equilibrium. But, the strategy pair $(s_2,s_2) = (Betray, Betray)$ is the unique pure strategy Nash equilibrium of the game. This steady-state solution corresponds to $T \rightarrow 0$ where we assume that the players are perfectly rational. To explore the probability of choosing each strategy at other temperatures we employed the formulation presented in Eq.~(\ref{eq:4}). Accordingly, Fig.~\ref{fig1}A1 shows the probability distribution of a player in the Prisoner's Dilemma over her strategies in terms of her temperature. These probabilities are similar for each player since irrespective of the other player's choice, $s_2 = Betray$ is the dominant strategy and, besides, we assumed that the players are characterized by identical temperatures ($T_1 = T_2 = T$). In this case, as depicted in Fig.~\ref{fig1}A2 $p(s_1)|_{T \rightarrow 0} = 0$ and $p(s_2)|_{T \rightarrow 0} = 1$ when the players are perfectly rational, whereas $p(s_1)|_{T \rightarrow \infty} = p(s_2)|_{T \rightarrow \infty} = \frac{1}{2}$ when the players are perfectly irrational. Furthermore, the joint probabilities of four possible outcomes are shown in Fig.~\ref{fig1}B1. In this case, $P(s_2,s_2)|_{T \rightarrow 0} = 1$ and other probabilities vanish at zero temperature, whereas at extremely high temperatures $P(s_i,s_j)|_{T \rightarrow \infty} = \frac{1}{4}$ for all possible outcomes, as illustrated in Fig.~\ref{fig1}B2.

\begin{table}
\centering
\newcommand{\noline}[1]{\multicolumn{1}{c}{#1}}
\caption{{\bf Examples of classic $2 \times 2$ strategic games.} (Left) Prisoner's Dilemma which has a unique pure strategy Nash equilibrium. (Middle) Battle of the Sexes which has two pure strategy Nash equilibria as well as a mixed strategy Nash equilibrium. (Right) Matching Pennies which has a unique mixed strategy Nash equilibrium. The two rows (columns) correspond to the two possible strategies of player 1 (player 2). The numbers in each box represent the players' payoffs to the strategy profile to which the box corresponds, with player 1's payoff listed first. The best response of each player is marked with an asterisk.}

\begin{tabular}[t]{c|c|c|}
\noline{} & \noline{\textit{Quiet}} & \noline{\textit{Betray}} \\
\cline{2-3}
\textit{Quiet} & ~2,2~ & ~0,$3^*$~ \\ \cline{2-3}
\textit{Betray} & $3^*$,0 & $1^*,1^*$ \\ \cline{2-3}
\end{tabular}
\hspace*{0.2cm}
\begin{tabular}[t]{c|c|c|}
\noline{} & \noline{\textit{Boxing}} & \noline{\textit{Shopping}} \\
\cline{2-3}
\textit{Boxing} & ~~$2^*,1^*$~~ & ~0,0~ \\ \cline{2-3}
\textit{Shopping} & 0,0 & $1^*,2^*$ \\ \cline{2-3}
\end{tabular}
\hspace*{0.2cm}
\begin{tabular}[t]{c|c|c|}
\noline{} & \noline{\textit{Head}} & \noline{\textit{Tail}} \\
\cline{2-3}
\textit{Head} & ~$1^*$,-1~ & ~-1,$1^*$~ \\ \cline{2-3}
\textit{Tail} & -1,$1^*$ & $1^*$,-1 \\ \cline{2-3}
\end{tabular}
\vspace{0.4cm}
\label{table1}
\end{table}

\begin{figure}[t!]
\centering
\includegraphics[scale = 1]{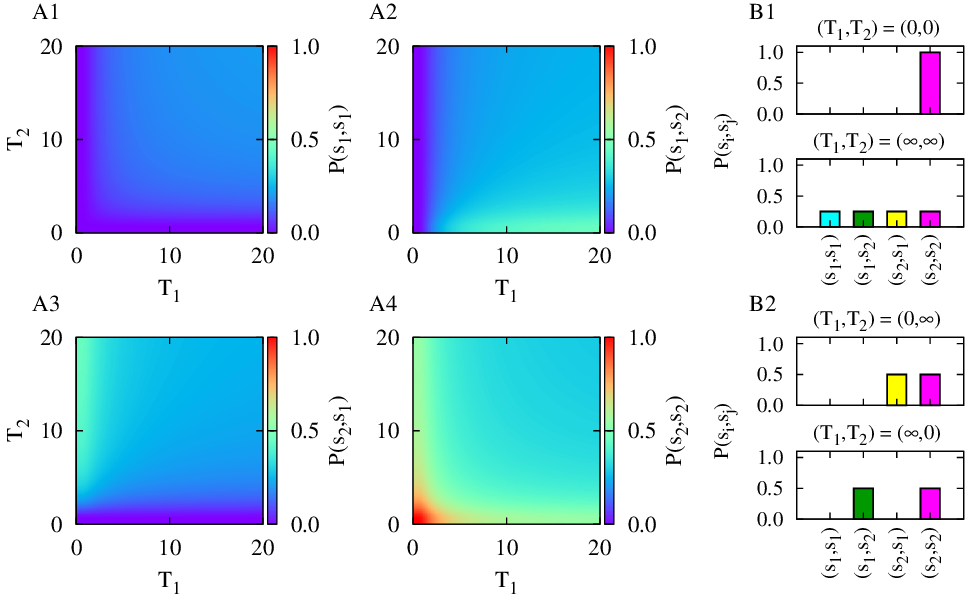}
\caption{{\bf Probabilities of possible outcomes of the Prisoner's Dilemma for non-identical temperatures.} (\textbf{A1-A4}) Joint probabilities of possible outcomes of the game for non-identical temperatures of the players with two strategies $s_1 = Quiet$ and $s_2 = Betray$. (\textbf{B1,B2}) Examplary joint probabilities of possible outcomes for asymptotic temperatures indicated by a temperature pair ($T_1,T_2$) above each panel.}
\label{fig2}
\end{figure}

To explore the effect of non-identical temperatures ($T_1 \neq T_2$) of the two players on the outcome of the game, the joint probabilities of possible outcomes are shown in Fig.~\ref{fig2}A1-A4. In this setting, each of the four possible outcomes are characterized by a wide range of the joint probabilities depending on the temperature of the players. In particular, rational choices are made near the $T_1 \rightarrow 0$ and $T_2 \rightarrow 0$ region highlighted by red color in Fig.~\ref{fig2}A4. For identical temperatures shown in Fig.~\ref{fig2}B1, i.e., $(T_1,T_2)=(0,0)$ and $(T_1,T_2)=(\infty,\infty)$, the results presented in Fig.~\ref{fig1}B2 are reproduced, i.e., $P(s_2,s_2)|_{T \rightarrow 0} = 1$ at zero temperatures, and $P(s_i,s_j)|_{T \rightarrow \infty} = \frac{1}{4}$ for all possible outcomes at extremely high temperatures. As an example for other temperatures, at $(T_1,T_2)=(0,\infty)$ and $(T_1,T_2)=(\infty,0)$ the two outcomes share a 50-50$\%$ chance of occurrence and other possibilities vanish, as shown in Fig.~\ref{fig2}B2.

\subsubsection{Non-Nash equilibrium solution concept}

Prisoner's Dilemma as presented in Table~\ref{table1} (left) is a special case which has a unique pure strategy Nash equilibrium, and the players are always confronted by the same utility difference, i.e., $\Delta U = \pm 1$ when switch between the two strategies. Therefore, the framework based on the Boltzmann weight presented in Eqs.~(\ref{eq:4})-(\ref{eq:7}) predicts deterministic choices at $T \rightarrow 0$ where the players are perfectly rational. However, perfect rationality in a wide range of games such as the Battle of the Sexes in Table~\ref{table1} (middle) and the Matching Pennies in Table~\ref{table1} (right) may imply probabilistic choices even at $T \rightarrow 0$, resulting in a mixed strategy Nash equilibrium. But, the Boltzmann weight formalism inherently lacks such an interpretation at $T \rightarrow 0$.

\begin{figure}[t!]
\centering
\includegraphics[scale = 1]{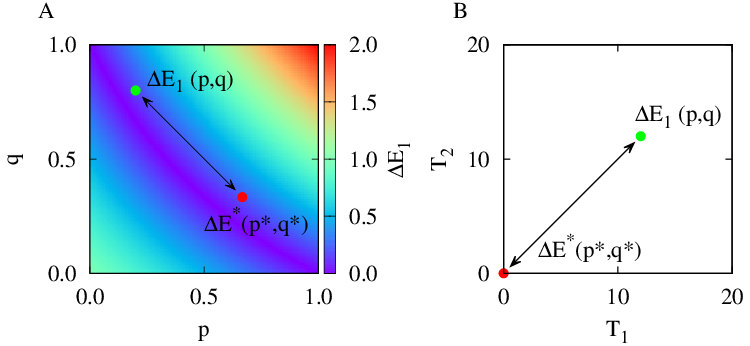}
\caption{{\bf Difference between the expected payoffs characterizes Nash equilibrium and non-Nash equilibrium states.} (\textbf{A}) Depending on the given joint probability $(p,q)$, the color-coded difference between the expected payoffs ($\Delta \mathcal{E}_1$; for player 1) of the two strategies determines Nash equilibrium (red circle) and non-Nash equilibrium (green circle) states, e.g., according to payoffs given in the Battle of the Sexes in Table~\ref{table1} (middle). (\textbf{B}) Each arbitrary state in the $p$-$q$ plane representing a $\Delta \mathcal{E}_1$ corresponds to a point in the $T_1$-$T_2$ plane which identifies the temperature pair ($T_1,T_2$) of the players in the game.}
\label{fig3}
\end{figure}

To resolve this issue, a modified framework should be able to (i) yield probabilistic choices at $T \rightarrow 0$, (ii) determine the player's temperature with respect to a given set of probabilities over strategies and, (iii) present a non-Nash equilibrium solution concept for predicting the probability of possible outcomes in a broad range of temperatures. Our framework emphasizes the difference of expected payoff  between the players' strategies rather than the corresponding utility difference. For instance, assume any two-player two-strategy game where player 1 (player 2) assigns probability $p$ ($q$) to her first strategy and probability $1 - p$ ($1 - q$) to her second strategy. Assume that the mixed strategy Nash equilibrium of the game is given as $(p^*,q^*)$. As illustrated in Fig.~\ref{fig3}A, e.g., for the Battle of the Sexes as shown in Table~\ref{table1} (middle), in the $p$-$q$ plane we define the difference between the expected payoffs (coded by color) of player 1 with respect to playing either of her two strategies, as follows:
\begin{equation}\label{eq:11}
{\Delta \mathcal{E}}_1(p,q) = \vert \mathcal{E}_1(s_1,q) - \mathcal{E}_1(s_2,1-q) \vert,
\end{equation}
where $\mathcal{E}_1(s_1,q)$ is her expected payoff when she uses the pure strategy that assigns probability 1 to $s_1$ and player 2 uses her mixed strategy $q$. By the same token, for player 2, $\Delta \mathcal{E}_2(p,q)$ can be written in a similar fashion.

At the mixed strategy Nash equilibrium, the difference between the expected payoffs is vanished, i.e., ${\Delta \mathcal{E}}^*(p^*,q^*) = {\Delta \mathcal{E}}^*(\frac{2}{3},\frac{1}{3}) = 0$ (Fig.~\ref{fig3}A, red circle). Any other combination of the joint probabilities is a non-Nash equilibrium state characterized by a non-vanishing difference between the expected payoffs, e.g., ${\Delta \mathcal{E}}_i(p,q) = {\Delta \mathcal{E}}_i(0.2,0.8) = 0.16$ (Fig.~\ref{fig3}A, green circle). Each point in the $p$-$q$ plane characterizing a ${\Delta \mathcal{E}}$ corresponds to a point in the $T_1$-$T_2$ plane which reveals the temperature of each player (as schematically shown in Fig.~\ref{fig3}B). However, this relationship may not be easy to extract due to a huge parameter space. Yet, one thing is for sure: The equilibrium state ${\Delta \mathcal{E}}^*(p^*,q^*)$ corresponds to the temperature pair $(T_1,T_2) = 0$ and any other state ${\Delta \mathcal{E}}(p,q)$ corresponds to an arbitrary (non-zero) temperature pair $(T_1,T_2)$ in Fig.~\ref{fig3}B.

To formulate this, the model must satisfy a few conditions. (i) Assuming that each player is characterized by a temperature that quantifies her rationality level, as the temperature changes from $T \rightarrow 0$ (perfect rationality) to $T \rightarrow \infty$ (perfect irrationality), the non-Nash equilibrium probabilities exhibit deviations from the Nash equilibrium probabilities. (ii) Asymptotically, at $T \rightarrow 0$ the non-Nash equilibrium solution must reproduce the Nash equilibrium probabilities (where ${\Delta \mathcal{E}}^*(p^*,q^*) = 0$), and at $T \rightarrow \infty$ strategies of the players are equally likely to be chosen. (iii) When the players are not perfectly rational, they strive to choose probabilities that minimize the difference between the expected payoffs of the two strategies. 

Inspired by the Boltzmann weight formalism introduced in Eqs.~(\ref{eq:4})-(\ref{eq:7}), one of the forms that such a function could be written is:
\begin{equation}\label{eq:12}
p_i(s_i,T_i) = p_i^*(s_j,0) + [C - p_i^*(s_i,0)] \, e^{- \Delta\mathcal{E}_i / kT_i},\\
\end{equation}
where $C$ guarantees that $0 \leq p_i(s_i,T_i) \leq 1$, $k = 1.0 \times 10^{-3}$ is a scaling factor, $p_i(s_i,T_i)$ denotes the probability with which the player $i$ with temperature $T_i$ may choose her strategy $s_i$, $p_i^*(s_i,0)$ is the Nash equilibrium probability of playing the strategy $s_i$ at $T \rightarrow 0$.

Assuming that the mixed strategy Nash equilibrium of the game is given, the predicted probabilities are not unique, rather $C = p_{\rm min} = 0$ results in $p_i(s_i,T_i) < p_i^*(s_i,0)$ and $C = p_{\rm max} = 1$ leads to $p^{\prime}_i(s_i,T_i) > p_i^*(s_i,0)$. This is because when the rationality of a player is decreased by increasing the temperature, the probability that she assigns to a strategy may be either smaller or greater than the probabilities assigned at the Nash equilibrium of the game.

\subsubsection{Prediction of temperature and non-Nash equilibrium probability}

Based on the aforementioned argumentation and formulation presented in Eq.~(\ref{eq:12}), given the Nash equilibrium of the game two scenarios can occur for the prediction of temperatures and non-Nash equilibrium probabilities:

\begin{itemize}

\item The joint probability $(p,q)$ of the players over their strategies is given and the question is to find temperatures $(T_1,T_2)$ of the players that resulted in such a joint probability.

\item Temperatures of the players are given and the question is to find the joint probability with which the players would choose their strategies.

\end{itemize}

\begin{figure}[t!]
\centering
\includegraphics[scale = 1]{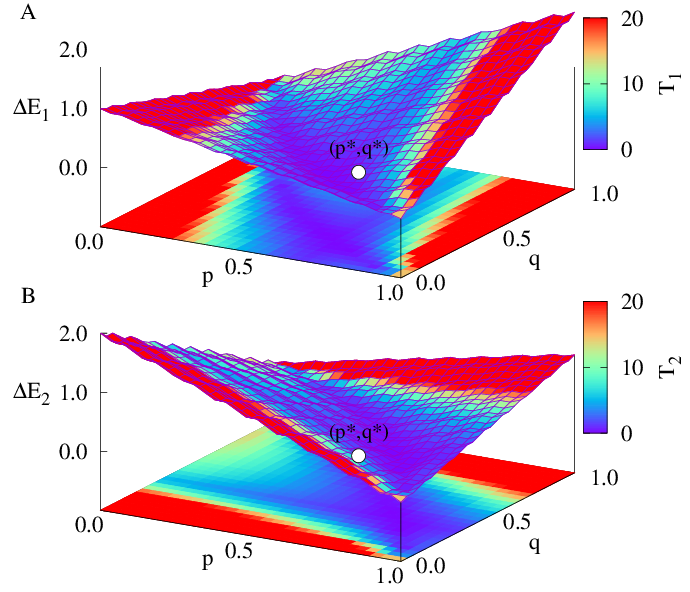}
\caption{{\bf Prediction of the temperature of the players.} The joint probability $(p,q)$ of the players is systematically varied and the resultant difference between the expected payoffs is calculated according to Eq.~(\ref{eq:11}) for player 1 (\textbf{A}) and player 2 (\textbf{B}). The temperature of the players is then calculated via Eq.~(\ref{eq:13}). The white circles indicate the Nash equilibrium joint probability $(p^*,q^*) = (\frac{2}{3},\frac{1}{3})$.}
\label{fig4}
\end{figure}

The results for the first scenario are shown in a 4-dimensional representation in Fig.~\ref{fig4} where the z-axis indicates ${\Delta \mathcal{E}_i}$ and the colorbar codes the temperature of the player $i$. The results are depicted for the Battle of the Sexes where the payoffs of players are given in Table~\ref{table1} (middle). In Fig.~\ref{fig4}, the joint probability $(p,q)$ of the players over their strategies is varied and the resultant difference between the expected payoffs is calculated for the two players according to Eq.~(\ref{eq:11}), i.e., player 1 (${\Delta \mathcal{E}_1}$ in Fig.~\ref{fig4}A) and player 2 (${\Delta \mathcal{E}_2}$ in Fig.~\ref{fig4}B). Then, the temperature of the player $i$ for a given set of choice probabilities can be calculated by rearranging Eq.~(\ref{eq:12}), as follows (temperature dependency of probabilities are omitted for simplicity):
\begin{equation}\label{eq:13}
T_i = 
\begin{cases}
- \dfrac{\Delta\mathcal{E}_i}{k \ln \left( \frac{p_i(s_i) - p_i^*(s_i)}{p_{\rm min} - p_i^*(s_i)} \right) },& {\rm if} ~ p_i(s_i) < p_i^*(s_i),\\
- \dfrac{\Delta\mathcal{E}_i}{k \ln \left( \frac{p_i(s_i) - p_i^*(s_i)}{p_{\rm max} - p_i^*(s_i)} \right)},& {\rm if} ~ p_i(s_i) > p_i^*(s_i),\\
0 ,& {\rm if} ~ p_i(s_i) = p_i^*(s_i),
\end{cases}
\end{equation}
where the resultant temperature somehow quantifies the player's rationality based on the deviation of the given probability ($p_i$) from the given Nash equilibrium probability ($p_i^*$) that is reflected in the difference of expected payoffs over each strategy (${\Delta \mathcal{E}_i}$).

As it is shown in Fig.~\ref{fig4}, cool (bluish) colors represent more rationally assigned probabilities near $T \rightarrow 0$. Specifically, the results presented in Fig.~\ref{fig4} confirm that in some sets of the joint probabilities a basin of attraction (bluish region) is formed near the Nash equilibrium marked by white circle, i.e., $(p^*,q^*) = (\frac{2}{3},\frac{1}{3})$, where the difference between the expected payoffs for each of the two strategies is minimum ($\Delta \mathcal{E}_i \approx 0$) and both players are characterized by low temperatures (bluish region). However, as the joint probability deviates from the Nash equilibrium, $\Delta \mathcal{E}_i$ and $T_i$ are both increased, implying irrational choices at the limit $T \rightarrow \infty$ (reddish region).

\begin{figure}[t!]
\centering
\includegraphics[scale = 1]{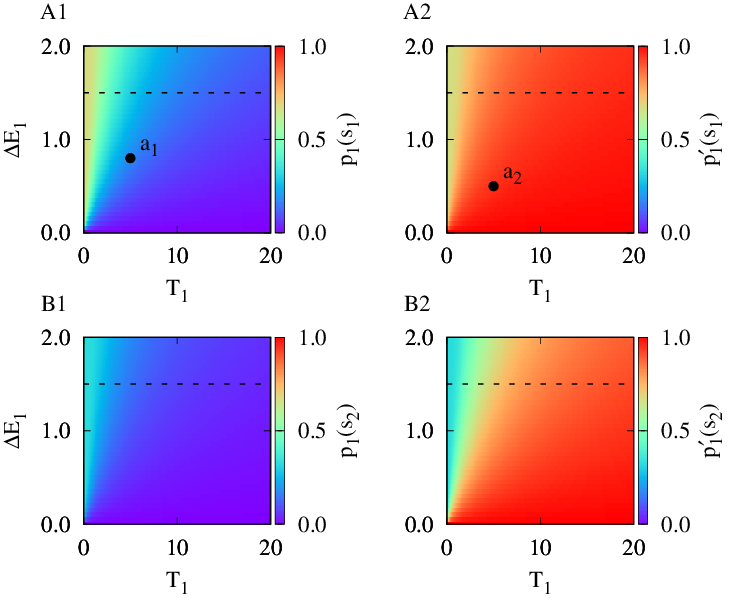}
\caption{{\bf Prediction of non-Nash equilibrium probabilities over strategies.} Color-coded non-Nash equilibrium probabilities over strategies of player 1 for payoffs given in the Battle of the Sexes in Table~\ref{table1} (middle) as a function of temperature and expected payoffs. These probabilities correspond to the conditions where the predicted non-Nash equilibrium probability is smaller (\textbf{A1,B1}) or greater (\textbf{A2,B2}) than the Nash equilibrium probability of the player based on Eq.~(\ref{eq:12}). The two points that are marked represent $\rm a_1$ = (5,0.8) and $\rm a_2$ = (5,0.5) in the $T$-$\Delta \mathcal{E}$ plane. Dashed horizontal lines represent $\Delta \mathcal{E}_1 = 1.5$ which was used to depict Fig.~\ref{fig6}.}
\label{fig5}
\end{figure}

The second scenario assumes that the temperatures of the players are given and the respective joint probability of the players is to be determined based on Eq.~(\ref{eq:12}). In this case, there are three pairs of parameters, i.e., $(p,q)$, $(\Delta \mathcal{E}_1,\Delta \mathcal{E}_2)$ and $(T_1,T_2)$, where only one of them (the temperature pair) is given. Therefore, we also varied the difference between the expected payoffs in order to visualize the results for each player. These results for player 1 are shown in Fig.~\ref{fig5} where the temperature of the player ($T_1$) and the difference of the expected payoffs ($\Delta \mathcal{E}_1$) are systematically varied and the resultant color-coded probability is calculated based on Eq.~(\ref{eq:12}). These results demonstrate that at a given temperature there are two sets of non-Nash equilibrium probabilities over the strategy profiles of the players (denoted by $p$ and $p^{\prime}$). In other words, the probability that a player assigns to a strategy can be either smaller (Fig.~\ref{fig5}A1 and B1) or greater (Fig.~\ref{fig5}A2 and B2) than the Nash equilibrium probability, as the temperature (irrationality) increases. The probabilities of player 2 can be calculated in the same way.

For instance, consider ($T,\Delta \mathcal{E}$) pairs at a constant temperature $T_1 = 5$ represented, e.g., by the $\rm a_1$ = (5,0.8) and $\rm a_2$ = (5,0.5) points marked in Fig.~\ref{fig5}A1 and A2. The corresponding probabilities are $p_1(s_1)|_{\rm a_1}= 0.10$ and $p_1^{\prime}(s_1)|_{\rm a_2}= 0.97$, respectively. Since there are two sets of probabilities, the fraction that these probabilities are played then can be simply approximated by the Boltzmann factor of their respective difference between the expected payoffs, i.e., $e^{(\Delta \mathcal{E}_1|_{\rm a_1} - \Delta \mathcal{E}_1|_{\rm a_2})/kT}$. This implies that when $\Delta \mathcal{E}_1|_{\rm a_1} \approx \Delta \mathcal{E}_1|_{\rm a_2}$, there is a 50-50$\%$ chance that each probability will be chosen. But when $\Delta \mathcal{E}_1|_{\rm a_1} \gg \Delta \mathcal{E}_1|_{\rm a_2}$ or $\Delta \mathcal{E}_1|_{\rm a_1} \ll \Delta \mathcal{E}_1|_{\rm a_2}$, the probability set with smaller $\Delta \mathcal{E}$ is more likely to be chosen by the player.

\begin{figure}[t!]
\centering
\includegraphics[scale = 1]{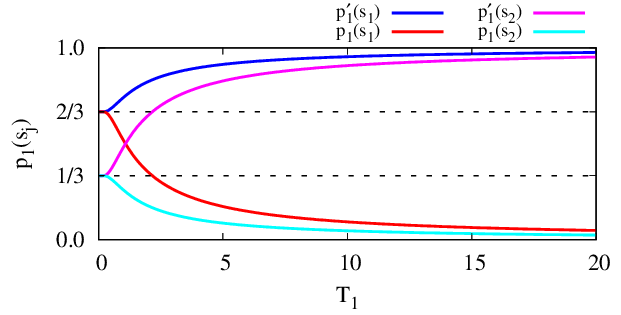}
\caption{{\bf Non-uniqueness of the predicted non-Nash equilibrium probabilities.} The predicted non-Nash equilibrium probabilities based on Eq.~(\ref{eq:12}) are not unique. They can be smaller (red and cyan) or greater (blue and magenta) than the respective Nash equilibrium probabilities. The figure is depicted by assuming that $\Delta \mathcal{E}_1 = 1.5$. Dashed horizontal lines represent Nash equilibrium probabilities at $T \rightarrow 0$ as a reference, i.e., $(p^*,q^*) = (\frac{2}{3},\frac{1}{3})$.}
\label{fig6}
\end{figure}

This non-uniqueness of predicted non-Nash equilibrium probabilities is the fundamental difference between decision theory and game theory. In decision theory, the choice of a player does not affect the probability with which the other player would choose a strategy. However, in game theory the players assign a probability to a strategy in an interactive manner. To further clarify this point, the predicted non-Nash equilibrium probabilities for player 1 are depicted for a specific value of the difference between the expected payoffs (e.g., $\Delta \mathcal{E}_1 = 1.5$) in Fig.~\ref{fig6}. For player 1, Nash equilibrium probabilities are $p_1^*(s_1) = \frac{2}{3}$ and $p_1^*(s_2) = \frac{1}{3}$ at $T \rightarrow 0$. As the temperature increases the players become more irrational and the predicted non-Nash equilibrium probabilities converge to their asymptotic values (i.e., 0 or 1) at $T \rightarrow \infty$. Meanwhile, these predicted non-Nash equilibrium probabilities can be smaller (Fig.~\ref{fig6}, red and cyan) or greater (Fig.~\ref{fig6}, blue and magenta) than the Nash equilibrium probabilities of the player. Interestingly, at $T \rightarrow \infty$, this yields two sets of probabilities for the strategy $s_1$ for instance, i.e., $p_1(s_1) = 0$ and $p_1^{\prime}(s_1) = 1$, both characterized by the same $\Delta \mathcal{E}_1 = 1.5$. This implies that these probabilities will be played with a 50-50$\%$ chance, as predicted before. Given the parameters, non-Nash equilibrium probabilities of player 2 can be calculated in a similar fashion.


\subsubsection{Relative entropy for probability distributions}

A crucial difference between the pure strategy and mixed strategy games is that in a pure strategy game the Nash equilibrium is achieved when the perfectly rational players (at $T \rightarrow 0$) make deterministic choices. In contrast, in a mixed strategy game the Nash equilibrium is achieved when the players make probabilistic choices even if they are perfectly rational (at $T \rightarrow 0$). In other words, in this case the perfect rationality assumption compels players to choose stochastically to maximize their expected payoff.

Interestingly, this is reflected in the effective information or relative entropy of probability distributions over possible outcomes at $T \rightarrow \infty$ (perfect irrationality with maximum entropy distribution on the states of the system) with respect to probability distributions over possible outcomes at $T \rightarrow 0$ (perfect rationality with minimum entropy distribution on the states of the system). Let's consider a simple example of two-player games such as the Prisoner's Dilemma as in Table~\ref{table1} (left) and the Battle of the Sexes as in Table~\ref{table1} (middle). At $T \rightarrow \infty$, in both games the players choose their strategies randomly with equal probability, i.e., $(p,q) = (\frac{1}{2},\frac{1}{2})$, resulting in $H_{\rm max}|_{T \rightarrow \infty} = 2.00$ bits according to Eq.~(\ref{eq:9}). However, at $T \rightarrow 0$ in the Prisoner's Dilemma the pure strategy Nash equilibrium reads $(p,q) = (0,0)$, resulting in $H_{\rm min}|_{T \rightarrow 0} = 0.00$ bits, whereas in the Battle of the Sexes the mixed strategy Nash equilibrium reads $(p,q) = (\frac{2}{3},\frac{1}{3})$, resulting in $H_{\rm min}|_{T \rightarrow 0} = 1.84$ bits. Therefore, the effective information can be calculated based on Eq.~(\ref{eq:8}), as follows
\begin{equation}\label{eq:14}
\begin{aligned}
I_{\rm PD} &= H_{\rm max}|_{T \rightarrow \infty} - H_{\rm min}|_{T \rightarrow 0} = 2.00 \, {\rm bits} - 0.00 \, {\rm bits} = 2.00 \, {\rm bits},\\
I_{\rm BoS} &= H_{\rm max}|_{T \rightarrow \infty} - H_{\rm min}|_{T \rightarrow 0} = 2.00 \, {\rm bits} - 1.84 \, {\rm bits} = 0.16 \, {\rm bits},
\end{aligned}
\end{equation}
where the Shannon entropies were calculated over all possible outcomes of the game. This example illustrates that a game with a pure strategy Nash equilibrium can be characterized by an effective information that is equal to its maximum entropy distribution on the states of the system at $T \rightarrow \infty$, i.e., $I_{\rm Pure} = H_{\rm max}|_{T \rightarrow \infty}$. But, a game with a mixed strategy Nash equilibrium can take effective information values less than the maximum entropy distribution, i.e., $I_{\rm Mixed} \leq H_{\rm max}|_{T \rightarrow \infty}$, due to a non-zero entropy of probability distribution at $T \rightarrow 0$.


\subsection{Analysis of the experimental dataset}
\subsubsection{Characteristics of the experimental dataset}

\begin{figure}[t!]
\centering
\includegraphics[scale = 1]{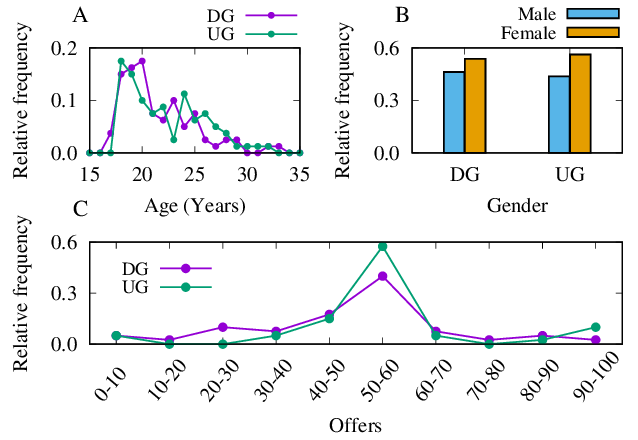}
\caption{{\bf Characteristics of the experimental dataset.} (\textbf{A}) The age distribution of players participating in the Dictator game (DG) and the Ultimatum game (UG). (\textbf{B}) The gender distribution of players participating in each game. (\textbf{C}) The relative frequency of the offered points (in the 0-100 range) in each game which are averaged over bins of size 10 points to enable a comparison between the outcomes of the two games.}
\label{fig7}
\end{figure}

We used previously published~\cite{wu2019gossip} experimental data to validate our model predictions. In the experiment (see Methods)~\cite{wu2019gossip}, 40 triads out of 80 triads each including two players (player 1 and player 2) were involved the Dictator game and 40 triads played the Ultimatum game. Analysis of the characteristics of the experimental dataset is shown in Fig.~\ref{fig7}. Particularly, the age (Fig.~\ref{fig7}A) and gender (Fig.~\ref{fig7}B) distributions of players participating in each game are fairly similar, suggesting that although the participants were randomly assigned as player 1 and 2, they share similar characteristics.

Based on the experimental paradigm, the payoff tables of Dictator game and Ultimatum game are presented in Table~\ref{table2}. As described in the Methods, in both games player 1 is endowed with 100 points and could offer arbitrary points in the range 0-100 to player 2. The difference between the two games is that the Dictator game captures a decision by the single player 1, i.e., to offer points to player 2 or not, and player 2 has no choice but to accept the offered amount (Table~\ref{table2}, top). In the Ultimatum game, however, player 2 has the option to accept or reject the offer. If the offer is accepted, the 100 points would be shared as proposed, otherwise, neither of the players receives anything (Table~\ref{table2}, bottom). As per the experimental setup~\cite{wu2019gossip}, we focus on a strategic situation in which actions in the Dictator game and the Ultimatum game are chosen once and for all (i.e., a one-shot game).

In this setting, the columns in Table~\ref{table2} represent the players' payoffs (with the player 1's payoff listed first) to the 101 possible strategies of player 1, indicating the amount of points she can offer in the Dictator game or the Ultimatum game, i.e., $\mathcal{S}_1 = \lbrace s_1,s_2,s_3,...,s_{101} \rbrace = \lbrace 0,1,2,...,100 \rbrace$, whereas the rows correspond to the possible strategies of player 2, i.e., $\mathcal{S}_2 = \lbrace s_1 \rbrace = \lbrace accept \rbrace$ in the Dictator game, and $\mathcal{S}_2 = \lbrace s_1,s_2 \rbrace = \lbrace accept,reject \rbrace$ in the Ultimatum game. Accordingly, the theory of rational choice predicts that the Dictator game has a unique pure strategy Nash equilibrium characterized by the strategy pair $(s_1,s_1) = (0,accept)$, with the player 1's strategy listed first. The Ultimatum game, on the other hand, has a unique pure strategy Nash equilibrium characterized by the strategy pair $(s_2,s_1) = (1,accept)$.

\begin{table}
\centering
\newcommand{\noline}[1]{\multicolumn{1}{c}{#1}}
\caption{{\bf The payoff tables of Dictator game and Ultimatum game.} The columns correspond to the 101 possible strategies of player 1 indicating the points she can offer, i.e., $\mathcal{S}_1 = \lbrace s_1,s_2,s_3,...,s_{101} \rbrace = \lbrace 0,1,2,...,100 \rbrace$, whereas the rows correspond to the possible strategies of player 2, i.e., $\mathcal{S}_2 = \lbrace s_1 \rbrace = \lbrace accept \rbrace$ in the Dictator game, and $\mathcal{S}_2 = \lbrace s_1,s_2 \rbrace = \lbrace accept,reject \rbrace$ in the Ultimatum game. The numbers in each box represent the players' payoffs to the strategy profile to which the box corresponds, with the player 1's payoff listed first. (Top) The payoff table of Dictator game which has a unique pure strategy Nash equilibrium characterized by the strategy pair $(s_1,s_1) = (0,accept)$ marked by asterisk. (Bottom) The payoff table of Ultimatum game which has a unique pure strategy Nash equilibrium characterized by the strategy pair $(s_2,s_1) = (1,accept)$ marked by asterisk.}

\begin{tabular}[t!]{c|c|c|c|c|c|c|c|}
\noline{} & \noline{$s_1$} & \noline{$s_2$} & \noline{$s_3$} & \noline{$s_{n-1}$} & \noline{$s_n$} & \noline{$s_{n+1}$} & \noline{$s_{101}$} \\
\cline{2-8}
$s_1$ & $100^*,0^*$ & 99,1 & 98,2 & ... & ... & ... & 0,100\\ \cline{2-8}
\end{tabular}\\
\vspace*{0.3cm}
\hspace*{0.0mm}
\begin{tabular}[t]{c|c|c|c|c|c|c|c|}
\noline{} & \noline{$s_1$} & \noline{$s_2$} & \noline{$s_3$} & \noline{$s_{n-1}$} & \noline{$s_n$} & \noline{$s_{n+1}$} & \noline{$s_{101}$} \\
\cline{2-8}
$s_1$ & 100,0 & $99^*,1^*$ & 98,2 & ... & ... & ... & 0,100\\ \cline{2-8}
$s_2$ & 0,0 & 0,0 & 0,0 & ... & ... & ... & 0,0\\ \cline{2-8}
\end{tabular}
\label{table2}
\end{table}

Since the participants in each triad playing the Dictator game or the Ultimatum game did not necessarily offer the same points in the range 0-100, we averaged the offered points in both of the experimental games over bins of size 10 points to be able to compare the outcomes of the two games (see Fig.~\ref{fig7}C). Consequently, we assumed that the strategy set of player 1 is reduced to 10 strategies (denoted by $\mathcal{B}_1$) characterized by the midpoint offers of each bin on average, i.e., $\mathcal{B}_1 = \lbrace b_1,b_2,b_3,...,b_{10} \rbrace = \lbrace 5,15,25,...,95 \rbrace$. The midpoint of each bin was calculated by averaging the upper and lower bounds of the bin interval.

Previous standard experimental results indicated that the mean offers in the Dictator game are around 30-40\% of the points endowed to player 1~\cite{bolton1998dictator,engel2011dictator}, whereas the mean (accepted) offers in the Ultimatum game are slightly greater, i.e., around 40-50\% of the endowed amount~\cite{guth1982experimental,roth1991bargaining,henrich2005economic}. The average offered points in the Dictator game (violet) and the Ultimatum game (green) datasets are shown in Fig.~\ref{fig7}C. Interestingly, only one participant playing the role of player 2 rejected the offer of player 1 in the Ultimatum game, therefore, we only focused on the behavior of player 1 in both games. Notably, the mean offered points at the bin 50-60 are relatively greater in the Ultimatum game (57.5\%) than the Dictator game (40\%), suggesting that when the responder (player 2) has the option to reject the offer, the proposers (player 1) seem to be more generous.


\subsubsection{Extracting rationality temperature from the dataset}

Following our model, it is possible to estimate the temperature of the players in a game based on their choices to approximate their level of rationality in decision making. In this framework, players who make more rational choices, e.g., offering very small amounts of points in the Dictator game and the Ultimatum game, are characterized by low temperatures ($T \rightarrow 0$), whereas players making irrational choices, i.e., offering very large amounts of points, are recognized by high temperatures ($T \rightarrow \infty$). Here, we sought to extract the temperature of players based on the amounts of points they offered in each experimental dataset for the Dictator game and the Ultimatum game.

\begin{figure}[t!]
\centering
\includegraphics[scale = 1]{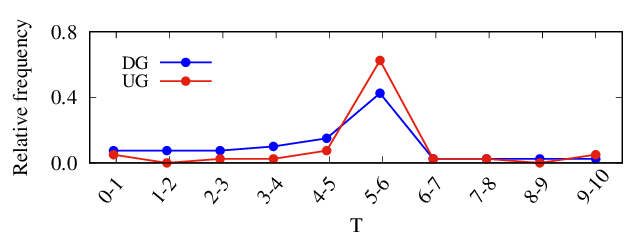}
\caption{{\bf Rationality temperature of the players.} Temperature distribution of the players participating in the Dictator game (DG) and the Ultimatum game (UG). The temperatures (in the 0-10 range) of the players in each game are calculated based on Eq.~(\ref{eq:13}) from the unbinned dataset and, then, averaged over bins of size of 1 units to enable a comparison between the temperatures of players in the two games.}
\label{fig8}
\end{figure}

The extracted temperatures of player 1 in each game are shown in Fig.~\ref{fig8}, which were calculated based on Eq.~(\ref{eq:13}). To do that, we assumed that the Nash equilibrium probabilities of possible outcomes for the unbinned offers are given as $p_1^*(s_1) = 1$ and $p_1^*(s_{-1}) = 0$ for the Dictator game, and $p_1^*(s_2) = 1$ and $p_1^*(s_{-2}) = 0$ for the Ultimatum game. Accordingly, the differences between the expected payoffs were determined by calculating the deviation of the expected payoff of an arbitrary non-Nash equilibrium choice ($s_i$) from the Nash equilibrium choice ($s^*$), i.e., $\Delta\mathcal{E}_1 = \vert \mathcal{E}_1(s_i) - \mathcal{E}_1(s^*) \vert$. These values were inserted into Eq.~(\ref{eq:13}) and the resultant temperature was calculated and, then, averaged over bins of size 1 units, as shown in Fig.~\ref{fig8}, identified by the midpoints of each bin, i.e., $\mathcal{T}_1 = \lbrace T_1,T_2,T_3,...,T_{10} \rbrace = \lbrace 0.5,1.5,2.5,...,9.5 \rbrace$.

The relative frequencies of temperatures in Fig.~\ref{fig8} were extracted from the relative frequencies of unbinned offers of the players in each game and, then, binned depending on which temperature interval they fell in. As shown in Fig.~\ref{fig8}, most of the players in each game (42.5\% in the Dictator game and 62.5\% in the Ultimatum game) are characterized by midrange values (i.e., bin 4-5) of temperature, implying that in a realistic situation only a small fraction of individuals are characterized by very low and very high temperatures. Therefore, only a small fraction of players might decide in a perfect rational (offering very small amounts of points) or irrational (offering very large amounts of points) manner.

\subsubsection{The model predicts experimental behavior}

According to the model, cool players are more likely to choose Nash or close-to-Nash equilibrium strategies (offering very small amounts of points) at $T \rightarrow 0$ to benefit more, whereas non-Nash equilibrium choices (offering very large amounts of points) are more likely to be chosen by hot players at $T \rightarrow \infty$, depending on the deviation of their expected payoff from the Nash equilibrium expected payoff. 

The participants in both experiments were all university students and shared similar age and gender characteristics, as shown in Fig.~\ref{fig7}A and B. Thus, it is not unreasonable to consider them as a small cohort (i.e., a group of individuals with shared characteristics) playing the Dictator game and the Ultimatum game. Therefore, the generosity of player 1 (i.e., the amount she offers) in each game can be considered as a \textit{test experiment} to estimate the temperature of players, which can be then used to predict the behavior of player 1 in the other game. For example, the rationality temperature extracted from the Dictator game can be used to predict the behavior of player 1 in the Ultimatum game and vice versa.

\begin{figure}[t!]
\centering
\includegraphics[scale = 1]{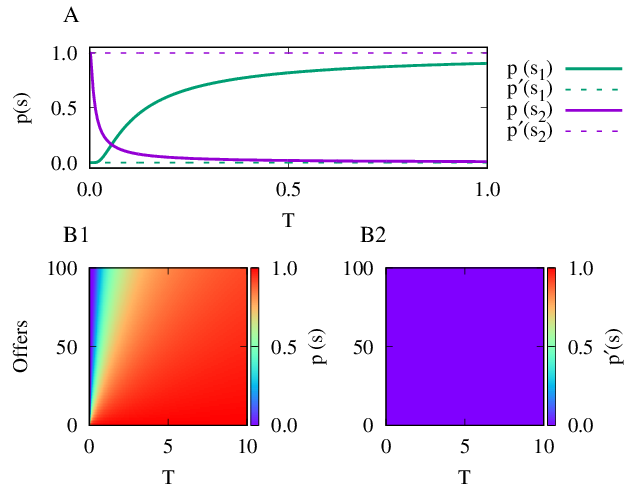}
\caption{{\bf Prediction of non-Nash equilibrium probabilities of possible outcomes in the Ultimatum game.} (\textbf{A}) An example of the predicted non-Nash equilibrium probabilities, $p(s_i)$ and $p^{\prime}(s_i)$, for the unbinned $s_1$ and $s_2$ strategies based on Eq.~(\ref{eq:12}). These predicted probabilities are not unique. They can be smaller or greater than the respective Nash equilibrium probabilities, i.e., $p^*(s_1) = 0$ and $p^*(s_2) = 1$. (\textbf{B1, B2}) Color-coded non-Nash equilibrium probabilities of offers of player 1 as a function of temperature.}
\label{fig9}
\end{figure}

In Fig.~\ref{fig9}, we used the estimated temperature distribution of the players in Dictator game to predict the probabilities of different offering choices in the Ultimatum game based on Eq.~(\ref{eq:12}). The reverse scenario reproduces qualitatively similar predictions as well. As an example, Fig.~\ref{fig9}A shows the predicted probabilities of choosing the unbinned strategies $s_1$ (violet) and $s_2$ (green). Here, we emphasize that the predicted probabilities of possible outcomes are not unique since once the Nash equilibrium probabilities are given, i.e., $p^*(s_1) = 0$ and $p^*(s_2) = 1$, the players may assign smaller or greater non-Nash equilibrium probabilities than the Nash equilibrium probabilities as the temperature increases. Therefore, Eq.~(\ref{eq:12}) yields two sets of probabilities indicated as $p$ (solid lines) and $p^{\prime}$ (dashed lines) in Fig.~\ref{fig9}A.

To explore the entire range of parameters, we systematically varied the temperature of players and the amount of offered points, and calculated both sets of probabilities in Fig.~\ref{fig9}B1 and B2. As shown in Fig.~\ref{fig9}B1, at low temperatures ($T \rightarrow 0$) the players tend to choose more rationally and avoid offering large amounts of points (blue region). However, as the temperature is increased the probability of choosing large offers is also increased such that at $T \rightarrow 10$ (equivalent to $T \rightarrow \infty$ here) all offering options are equally likely to occur (red region), i.e., $p(s_i) = 1 / n$, where $n$ is the possible number of offering strategies in the game. In this limit, the players are perfectly irrational and make decisions randomly. On the other hand, Fig.~\ref{fig9}B2 shows the other set of probabilities which are already at the minimum ($p^{\prime}(s_1) = 0$) or maximum ($p^{\prime}(s_2) = 1$) of their allowed range irrespective of the temperature.

\begin{figure}[t!]
\centering
\includegraphics[scale = 1]{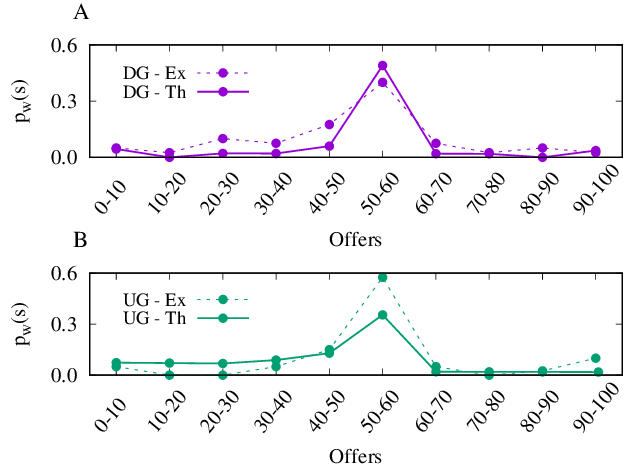}
\caption{{\bf Reproduction of experimental offers based on the model predictions.} (\textbf{A}) The weighted probability, $p_w(s)$, with which player 1 may offer points in the Dictator game (DG), predicted by the model (solid line) based on Eq.~(\ref{eq:15}). (\textbf{B}) The same probability (solid line) in the Ultimatum game (UG). Dashed lines in each panel indicate the experimental behavior of players as a reference.}
\label{fig10}
\end{figure}

We used the predicted non-Nash equilibrium probabilities of possible outcomes to test to what extent the experimentally observed behavior of players (see Fig.~\ref{fig7}C) in the Dictator game and the Ultimatum game can be reproduced by the model. This is shown in Fig.~\ref{fig10} where the rationality temperature of players (see Fig.~\ref{fig8}) extracted from the Ultimatum game was used to predict the behavior of player 1 in the Dictator game (Fig.~\ref{fig10}A, solid line) and vice versa (Fig.~\ref{fig10}B, solid line). However, according to the predicted non-Nash equilibrium probabilities in Fig.~\ref{fig9}B1, for player 1 at any arbitrary temperature there is a non-zero probability that she may choose a specific strategy. Therefore, the probability with which player 1 may choose the strategy $s_i$ is given by the weighted average of these probabilities, $p_w(s_i)$, according to the temperature distribution of players in each game, as follows:
\begin{equation}\label{eq:15}
p_w(s_i) = \frac{1}{n} \sum_{i = 1}^n w(T_i) \, p(s_i,T_i),
\end{equation}
where $n$ is the possible number of offering strategies in the game, $w(T_i)$ represents the likelihood of the temperature $T_i$ based on Fig.~\ref{fig8}, and $p(s_i,T_i)$ indicates the predicted non-Nash equilibrium probabilities of the strategy $s_i$ at the temperature $T_i$ according to Fig.~\ref{fig9}B1.

As shown in Fig.~\ref{fig10}, the offers predicted by the model (solid lines) are in good agreement with the experimentally observed behavior of players (dashed lines) both in the Dictator game (Fig.~\ref{fig10}A) and the Ultimatum game (Fig.~\ref{fig10}B). Since the players in Ultimatum game offered relatively greater points on average than the players in Dictator game (cf. Fig.~\ref{fig7}C, green and violet) and, therefore, were characterized by higher temperatures (cf. Fig.~\ref{fig8}, red and blue), the model-predicted offers of the players in Dictator game (Fig.~\ref{fig10}A), which were calculated based on the extracted temperatures of players in the Ultimatum game, are overestimated. On the contrary, the model-predicted offers of the players in Ultimatum game (Fig.~\ref{fig10}B) are underestimated. Nonetheless, in either case the distributions of offers predicted by the model (solid lines) fairly follow the offering behaviors that were observed experimentally (dashed lines).


\section{Discussion}

Inconsistency between the predictions of normative game theory and the results of real experiments has been a matter of debate for decades. Bounded rationality of players in a realistic situation has been recognized as one of the main sources of this discrepancy. To resolve this issue, decisions of the bounded rational players were modeled by choice probabilities which are given by the Boltzmann distribution~\cite{wolpert2006information,ortega2013thermodynamics,holehouse2022non}. Here, we argued that assigning a temperature to each player which mimics their level of sanity is able to parameterize this bounded rationality. By introducing the temperature parameter, we presented a simple and generic solution concept for predicting non-Nash equilibrium probabilities of possible outcomes in two-player strategic games based on the Boltzmann weight formalism. In this setting, temperature variations allow for two asymptotic behaviors, i.e., utility maximization by perfectly rational players at $T \rightarrow 0$ and equiprobable choices by perfectly irrational players at $T \rightarrow \infty$. Given the Nash equilibrium of the game and the probability distribution of the players over their strategies, our framework determines the temperature of the players and vice versa.

However, a practical, operational definition is required for estimating the human temperature (i.e., rationality) in order to calculate the probability distributions to cast some quantitative predictions. While without any prior knowledge it is impossible to independently estimate rationality either theoretically or empirically, the temperature of players can be, in principle, estimated by a test experiment. One approach to estimating or inferring the level of rationality without knowing the probabilities is through the use of observed choices and outcomes in experimental settings, where participants play games that simulate real-world strategic interactions. By analyzing the patterns of actual decisions made by the players in the test experiment and their deviations from the predictions of purely rational models, it would be possible to reverse-engineer the underlying level of rationality that led to those choices, generating a calibrated thermometer. With a reasonable assumption that game-theoretical experiments usually target a specific cohort of people, the temperature distribution extracted from the test experiment can be then used for the prediction of probability distributions in other games. 

To validate the ability of the presented model to explain non-Nash equilibrium strategic behavior in experimental games, we extracted temperature distribution of players from the experimental Dictator game and Ultimatum game. We used the rationality temperature extracted from the cohort playing Dictator game to predict the behavior of players in the Ultimatum game and vice versa. Our results confirmed that the distributions of offers predicted by the model are in good agreement with the experimental offers in each game. Therefore, in the context of our framework a practical way to measure the temperature of players can be outlined as follows:

\begin{itemize}

\item Conducting a simple test experiment (i.e., a game) to extract the deviations of players from the predictions of purely rational models.
\item Estimating the underlying temperature distribution of the players based on their choices and the outcome of the game.
\item Using the estimated, experimentally-grounded temperature distribution to predict probability distributions in other games by employing the model.

\end{itemize}

Previous attempts to model bounded rationality have several limitations and are still far from a complete understanding. In fact, the existing models such as the quantal response equilibrium~\cite{mckelvey1995quantal,mckelvey1998quantal}, epsilon equilibrium~\cite{radner1980collusive} as well as strategic thinking concepts like level-k reasoning or cognitive hierarchy models~\cite{nagel1995unraveling,camerer2004cognitive,costa2006cognition,evans2024bounded} fail in several occasions:

\begin{itemize}

\item They produce a set of possible outcomes, not a full distribution over all possible outcomes.
\item They cannot predict all possible joint mixed strategies chosen by the players.
\item They fail to predict the relative probabilities of the joint strategies in a game.
\item They lack the interpretation of playing probabilistically with mixed strategies even at $T \rightarrow 0$ where the players are perfectly rational.
\item When there are multiple pure or mixed strategy Nash equilibria in a game, they fail to assign relative probabilities to different solutions.

\end{itemize}
Here, we presented a plausible scenario to resolve these shortcomings by introducing a simple temperature-based paradigm which relies on the expected payoffs of the players for predicting non-Nash equilibrium probabilities of strategic games. Our approach offers several resolutions compared to the previous efforts:

\begin{itemize}

\item It provides a full distribution over all possible outcomes in a game as well as their probabilities.
\item It predicts all possible joint pure or mixed strategies in a game, even those with zero probability.
\item It relies on the expected payoffs of the players, enabling to predict the relative probabilities of the joint pure or mixed strategies.
\item It reinterprets rationality at $T \rightarrow 0$, offering a solution concept both for games with pure and mixed strategy Nash equilibria.
\item It determines the relative probabilities for multiple pure or mixed strategy Nash equilibria in a game.

\end{itemize}

Here, for simplicity and analytical tractability we restricted our analysis to classic two-player two-strategy games such as the Prisoner's Dilemma, Battle of the Sexes and Matching Pennies, as well as other games like the Dictator game and Ultimatum game. Nonetheless, our approach can be conceptually and numerically applied to two-player three-strategy games such as the Rock, Paper, Scissors as well as to various $N$-player $n$-strategy strategic games. However, translation of the results to the multi-player games with a wide range of strategies would require numerical methods and extensive computer simulations to search a huge parameter space and find the non-Nash equilibrium solutions. This remains to be appropriately addressed in the future work.

The Shannon entropy quantifies the degree of unpredictability or randomness in a probability distribution (here, over the players' choices)~\cite{wolpert2006information}. The use of temperature as a measure of the players' sanity allows us to draw an analogy between the concepts of entropy and temperature in thermodynamics. Just as higher temperatures in thermodynamics correspond to greater disorder or randomness, higher temperatures here correspond to greater randomness or uncertainty in the players' decisions. When the temperature is high, the (hot) players are more likely to choose strategies randomly, resulting in a mixed strategy equilibrium with higher entropy of probabilities. As the temperature decreases, cold players become more rational, resulting in deterministic choices and a pure strategy equilibrium with lower entropy.

Yet, even at $T \rightarrow 0$ when the players are perfectly rational, the outcome of the game can be probabilistic, e.g., in a game with mixed strategy Nash equilibrium such as the Battle of the Sexes and the Matching Pennies, i.e., in this case rationality means to play probabilistically. The Boltzmann weight in its classical form inherently lacks such an interpretation at $T \rightarrow 0$. Borrowing from the Boltzmann weight and by minimizing the difference between the expected payoff of the players with respect to playing either of their strategies, our generic framework was able to address probabilistic choices of players even at $T \rightarrow 0$.

In thermodynamics and statistical physics, the Boltzmann distribution gives the probability that a physical system at thermal equilibrium will be in a specific state based on the energy associated with that state. In the context of game theory, players are individuals or entities that make decisions and interact with one another in strategic situations. The Boltzmann weight provides a framework for evaluating the probability distribution of strategies adopted by these players that is influenced by the payoffs associated with different strategies as well as the rationality of the players. Living agents, such as human players in a game, often exhibit complex decision-making processes influenced by psychological and cognitive factors~\cite{roth1991bargaining,nagel1995unraveling,camerer2004cognitive,henrich2005economic}. The Boltzmann weight can be used to model the rationality of these agents and their propensity to adopt different strategies based on their perceived payoffs. This is particularly relevant in scenarios where individuals or groups are engaged in competitive or cooperative interactions, such as in economic competitions or social dilemmas~\cite{helbing1993stochastic,szabo2018social}. Non-living agents, on the other hand, include artificial intelligence algorithms, automated systems, or even simple decision-making processes. The Boltzmann weight provides a means to analyze the behavior of these non-living agents by assigning probabilities to different actions or responses, allowing to assess how these non-living agents might interact with their environment or with other agents in a game-theoretic context.

In the context of evolutionary dynamics of biological systems, for instance, the presence of uncertainty in the decision making of living organisms may shift a population away from the equilibrium point~\cite{nowak2004evolutionary,bauer2019stabilization}, making it difficult to predict behavior, but on the other hand, it creates diversity in the world, which can provide the raw material for natural selection~\cite{nowak2004emergence,traulsen2006evolution}. In fact, the effect of mutations on the behavior of organisms can be considered equivalent to changing the temperature in the decision-making process of that organism and creating new diversity in the population.

Since in the real world the temperature of the players in a game can be different, it is not possible for each player to make a correct decision in order to maximize her benefit without knowing the temperature of the other player.  Due to this uncertainty in a two-player game, the exact result of one round of game cannot be predicted, but the average results of repeating a large number of games between two players or in a population of people of a society or a species of organisms can be talked about.  In fact, we are faced with two types of uncertainty or non-Nash equilibrium states: One is the uncertainty in a single round of game caused by the temperature variability or irrationality of the players, and the other one is the deviation of the result of each specific game from the average results.

Ultimately, whether applied to human players in economic games, artificial intelligence algorithms in strategic environments, or evolutionary dynamics within biological populations, the Boltzmann weight enriches our understanding of agent behavior and its implications for strategic outcomes. Yet the Boltzmann distribution cannot be applied to systems out of equilibrium. Alternatively, by reinterpreting the concept of temperature, our framework represents a paradigm to predict non-Nash equilibrium probabilities of outcomes in a strategic game and offers a resolution for discrepancies between the predictions of normative game theory and the results of real experiments.


%


%


\section*{CRediT Author Statement}

\textbf{Mojtaba Madadi Asl:} Methodology, Formal analysis, Investigation, Visualization, Writing - original draft, Writing - review \& editing, Project administration. \textbf{Mehdi Sadeghi:} Conceptualization, Methodology, Formal analysis, Writing - review \& editing, Supervision.


\section*{Declaration of Competing Interests}

The authors declare that they have no known competing financial interests or personal relationships that could have appeared to influence the work reported in this paper.



\section*{Data availability}
Previously published~\cite{wu2019gossip} experimental data were used to validate model predictions. The experimental dataset analyzed during this study is available at \href{https://osf.io/ft8mu/}{https://osf.io/ft8mu/}.


\section*{Code availability}
All other data used to produce the figures were generated via numerical simulations. The simulation code is publicly accessible at \href{https://github.com/MMadadiAsl/Non-Nash-equilibrium-probability}{https://github.com/MMadadiAsl/Non-Nash-equilibrium-probability}.


%



\bibliography{references}
\bibliographystyle{vancouver}
\addcontentsline{toc}{section}{References}

\end{document}